\documentstyle[12pt,epsfig,twoside]{article}

\newcommand{\mev}{{\mathrm{\,Me\kern -0.1em V}}}
\newcommand{\mevc}{{{\mathrm{\,Me\kern -0.1em V\!/}c}}}
\newcommand{\mevcc}{{{\mathrm{\,Me\kern -0.1em V\!/}c^2}}}
\newcommand{\gev}{{\mathrm{\,Ge\kern -0.1em V}}}
\newcommand{\gevc}{{{\mathrm{\,Ge\kern -0.1em V\!/}c}}}
\newcommand{\gevcc}{{{\mathrm{\,Ge\kern -0.1em V\!/}c^2}}}

\def\eLik{{\cal L}}
\def\Ns{{\rm N}_{\rm s}}
\def\f{{\rm f}}
\def\F{{\rm F}}

\def\d{{\rm d}}
\def\P{{\cal P}}
\def\seP{{_{es}{\cal P}}}
\def\Mtrue{{\hbox{\bf{M}}}}
\def\mes{{m_{\rm ES}}}

\def\nast{{\rm n}}

\newcommand{\deltaE}{\Delta E}

\newcommand{\beq}{\begin{equation}}
\newcommand{\eeq}{\end{equation}}
\newcommand{\beqn}{\begin{eqnarray}}
\newcommand{\eeqn}{\end{eqnarray}}

\def\su3bris{SU(\slashchar{3})}

\def\L02{{\{L_0,L_2\}}}

\def\inPlots{_{\rm in}{\cal P}lots}
\def\inPlot{_{\rm in}{\cal P}lot}
\def\sPlots{\hbox{$_s$}{\cal P}lots}
\def\sPlot{\hbox{$_s$}{\cal P}lot}
\def\P{{\cal P}}
\def\sP{{_s{\cal P}}}
\def\Fisher{{\cal F}}
\def\M{{\rm M}}
\def\V{\hbox{\bf V}}
\def\Deltax{{\delta x}}
\def\chiant{e \subset \Deltax}
\def\Nx{N_{\Deltax}}

\epsfverbosetrue
\begin{document}

\vspace{1.cm}

\begin{center}
{\LARGE\bf $\sPlot:$

\vspace{0.5cm}
a statistical tool to unfold data distributions
}

\vskip 1.truecm
\centerline{\bf \large M.~Pivk$^{\,a}$ and F.R.~Le~Diberder$^{\,b}$}
\vspace{0.8cm}
{\small \em $^{a}$ CERN, \\
CH-1211 Geneva 23, Switzerland}\\[0.2cm]

{\small \em $^{b}$ Laboratoire de l'Acc\'el\'erateur Lin\'eaire,\\
                   IN2P3-CNRS et Universit\'e de Paris-Sud, 
                   F-91898 Orsay, France}

\end{center}
\vspace{1.8cm}

{\abstract{\it 
The paper advocates the use of a statistical tool dedicated to the exploration of data samples 
populated by several sources of events.
This new technique, called $\sPlot$, is able to unfold the contributions of the different sources to the distribution of a data sample in a given variable. The $\sPlot$ tool applies in the context of a Likelihood fit which is performed on the data sample to determine the yields of the various sources.}}% It is assumed that the fit relies on discriminating variables uncorrelated to the variable under consideration.}}

\newpage
\tableofcontents

\newpage

\section{Introduction}
This paper describes a new technique to explore a data sample when the later consists of several sources of events merged into a single sample of events. The events are assumed to be characterized by a set of variables which can be split into two components. The first component is a set of variables for which the distributions of all the sources of events are known: below, these variables are collectively referred to as a (unique) {\em discriminating} variable. The second component is a set of variables for which the distributions of some sources of events are either truly unknown or considered as such: below, these variables are collectively referred to as a (unique) {\em control} variable.

The new technique, termed $\sPlot$, allows to reconstruct the distributions for the control variable, independently for each of the various sources of events, without making use of any {\em a priori} knowledge on this variable. The aim is thus to use the knowledge available for the discriminating variable to be able to infer the behavior of the individual sources of events with respect to the control variable. An essential assumption for the $\sPlot$ technique to apply is that the control variable is uncorrelated with the discriminating variable.
 
The $\sPlot$ technique is developed in the context of a data sample analyzed using a maximum Likelihood method making use of the discriminating variable. Section~\ref{sec:basics} is dedicated to the definition of fundamental objects necessary for the following. Section~\ref{sec:inPlot} presents an intermediate technique, simpler but inadequate, which is a first step towards the $\sPlot$ technique. Section~\ref{sec:sPlot} is the core of the document where the $\sPlot$ formalism is developed (Section~\ref{sec:sPlotformalism}) and its properties explained in detail (Section~\ref{sec:sPlotprop}). Section~\ref{sec:Cuisine} then gives instructions about how to implement and use $\sPlot$. Finally, illustrations of $\sPlots$ are provided with simulated events (Section~\ref{sec:Illustrations}) and an application for branching ratios measurements (Section~\ref{sec:DBR}) is briefly described.

To provide some intuitive understanding of how and why the $\sPlot$ formalism works, the problem of reconstructing the true distributions is reconsidered in Appendix~\ref{sec:pedagogicalexamples}, in a simpler analysis framework. An extension of the $\sPlot$ technique is presented in Appendix~\ref{sectionextendedsplots}.

\section{Basics and definitions}
\label{sec:basics}
A common method used to extract parameters from a data sample is the maximum Likelihood method
which is briefly reviewed in Section~\ref{sec:Lik} since it constitutes the foundation of the $\sPlot$ technique. Section~\ref{sec:check} discusses the need for checks of an analysis based on the Likelihood method and introduces more precisely the goal of the $\sPlot$ technique.

\subsection{Likelihood method}
\label{sec:Lik}
One considers an extended Likelihood analysis of a data sample in which are merged several species of events. These species represent various signal components (ie. sources of events in which one is interested)
and background components (ie. irrelevant sources of events accompagnying the signal components) which all together account for the data sample. The log-Likelihood is expressed as:
\begin{equation}
\label{eq:eLik}
\eLik=\sum_{e=1}^{N}\ln \Big\{ \sum_{i=1}^{\Ns}N_i\f_i(y_e) \Big\} -\sum_{i=1}^{\Ns}N_i ~,
\end{equation}
where 
\begin{itemize}
\item $N$ is the total number of events in the data sample,
\item $\Ns$ is the number of species of events populating the data sample,
\item $N_i$ is the number of events expected on the average for the $i^{\rm th}$ species,
\item $y$ is the set of discriminating variables,
\item $\f_i$ is the Probability Density Function (PDF) of the discriminating variables for the $i^{\rm th}$ species,
\item $\f_i(y_e)$ denotes the value taken by the PDFs $\f_i$ for event $e$, 
the later being associated with a set of values $y_e$ for the set of discriminating variables,
\item $x$ is the set of control variables which, by definition, do not appear in the above expression of $\eLik$.
\end{itemize}
The log-Likelihood $\eLik$ is a function of the $\Ns$ yields $N_i$ and,
possibly, of implicit free parameters designed to tune the PDFs on the data sample.   
These parameters as well as the yields $N_i$ are determined by maximizing the above log-Likelihood.

\subsection{Analysis Validation }
\label{sec:check}
The crucial point for such an analysis of the data sample to be reliable
is to use an exhaustive list of sources of events combined with an accurate description of all the PDFs~$\f_i$. 
%Beside using accurate Monte-Carlo events checking the distribution shapes with the data is the best way.

To assess the quality of the fit,
one may rely on an evaluation of the goodness of fit based on the actual value obtained for the maximum of $\eLik$, but this is rarely convincing enough. A complementary quality check is to explore further the data sample by examining the distributions of control variables.
If the distributions of these control variables are known for at least one of the sources of events,
one can compare the expected distribution for this source to the one extracted from the data sample. 
In order to do so,
one must be able to unfold from the distribution of the whole data sample, 
the contribution arising from the source under scrutiny.

In some instances of control variables, 
the PDF might even be known for all the sources of events.
Such a control variable can be obtained for instance by removing one of the discriminating variables from the set~$y$ before performing again the maximum Likelihood fit,  
and considering the removed variable as a control variable~$x$. 
Another example is provided by a discriminating variable for which 
the distributions are known for all sources of events,
but which does not improve significantly the accuracy fo the fit, 
and is not incorporated in the set~$y$, for the sake of simplicity.

In an attempt to have access to the distributions of control variables, a common method consists in applying cuts which are designed to enhance the contributions to the data sample of particular sources of events (typically of signal species). Having enforced this enhancement, the distribution of~$x$ for the reduced data sample can be used to probe the quality of the fit through a comparison with a Monte Carlo simulated distribution. However, the result is frequently unsatisfactory: firstly because it can be used only if the signal has prominent features to be distinguished from the background, and secondly because of the cuts applied, a sizeable fraction of signal events can be lost, while a large fraction of background events may remain. Therefore, the resulting data distribution concerns a reduced subsample for which statistical fluctuations, or true anomalies, cannot be attributed unambiguously, neither to the signal, nor to the background. For example, one can be tempted to misinterpret an anomaly in the distribution of~$x$ coming from the signal as a harmless background fluctuation.

The aim of the $\sPlot$ formalism developed in this paper is to provide a convenient method to unfold the overall distribution of a mixed sample of events in a control variable~$x$ into the sub-distributions of the various species which compose the sample. It is a statistical technique which allows to keep all signal events while getting rid of all background events, and keeping track of the statistical uncertainties per bin.

More formally, one is interested in the true distribution (denoted in boldface $\Mtrue_\nast(x)$) of a control variable~$x$ for events of the $\nast^{\rm th}$ species, the later being any one of the $\Ns$ signal and background species. The purpose of this paper is to demonstrate that one can reconstruct $\Mtrue_\nast(x)$ from the sole knowledge of the PDFs of the discriminating variables $\f_i$, the first step being to proceed to the maximum Likelihood fit to extract the yields $N_i$. 

As an introduction, in Section~\ref{sec:inPlot}, the case is considered where the  variable~$x$ actually belongs to the set of~$y$ discriminating variables. That is to say that one makes the assumption opposite to the interesting one: $x$ is assumed to be totally correlated with~$y$.
Because of this total correlation, there exists a function of the~$y$ parameters which fully determines the 'control' variable, $x=x(y)$. In that case, while performing the fit, an {\it a priori} knowledge of the $x$-distributions is implicitly used, thus~$x$ cannot play the role of a control variable. Although the technique presented in the following Section is inadequate, it provides a natural first step towards $\sPlot$.

Section~\ref{sec:sPlot}, dedicated to the $\sPlot$ formalism, treats the interesting case, where~$x$ is truly a control variable uncorrelated with~$y$. In that case, while performing the fit, no {\it a priori} knowledge of the $x$-distributions is used.

\section{First step towards $\sPlot$: $\inPlot$} 
\label{sec:inPlot}
In this Section, one is considering a variable~$x$ which can be expressed as a function of the discriminating variables~$y$ used in the fit. A fit having been performed to determine the yields $N_i$ for all species, from the knowledge of the PDFs $\f_i$ and of the values of the $N_i$, one can define naively, for all events, the weight~\footnote{It was pointed out to the authors that a weight similar to the naive one of Eq.~(\ref{eq:weightxiny}) was introduced long ago in~\cite{bib:CondonCowell}.}
\begin{equation}
\label{eq:weightxiny}
\P_\nast(y_e)={N_\nast\f_\nast(y_e)\over\sum_{k=1}^{\Ns}N_k\f_k(y_e) } ~,
\end{equation}
which can be used to build the $x$-distribution $\tilde\M_\nast$ defined by:
\begin{equation}
\label{eq:inPlots}
N_\nast\tilde\M_\nast(\bar x)\Deltax~\equiv ~\sum_{\chiant} \P_\nast(y_e) ~,
\end{equation}
where the sum $\sum_{\chiant}$ runs over the $\Nx$ events 
for which $x_e$ (i.e. the value taken by the variable~$x$ for event $e$) lies in the $x$-bin centered on $\bar x$ and of total width $\Deltax$. 

In other words, $N_\nast\tilde\M_\nast(\bar x)\Deltax$ is the $x$-distribution obtained by histogramming events, using the weight of Eq.~(\ref{eq:weightxiny}). 

This procedure reproduces, on average, the true distribution $\Mtrue_\nast(x)$.
In effect, on average, one can replace the sum in Eq.~(\ref{eq:inPlots}) by the integral
\begin{equation}
\label{eq:sumGOTOintegral}
\left<\sum_{\chiant}\right>
\longrightarrow
\int\d y\sum_{j=1}^{\Ns}N_j\f_j(y)\delta(x(y)-\bar x)\Deltax ~.
\end{equation}
Similarly, 
identifying the number of events $N_i$ as determined by the fit to be the expected number of events, 
one obtains:
\begin{eqnarray}
\left<
N_\nast\tilde\M_\nast(\bar x)\right>
&=&\int\d y\sum_{j=1}^{\Ns}N_j\f_j(y)\delta(x(y)-\bar x)\P_\nast(y)
\nonumber \\
&=&\int\d y\sum_{j=1}^{\Ns}N_j\f_j(y)\delta(x(y)-\bar x){N_\nast\f_\nast(y)\over\sum_{k=1}^{\Ns}N_k\f_k(y) }
\nonumber \\
&=&N_\nast\int\d y\delta(x(y)-\bar x)\f_\nast(y)
\nonumber \\
\label{eq:naiveweightworks}
&\equiv&N_\nast\Mtrue_\nast(\bar x) ~.
\end{eqnarray}
Therefore, the sum over events of the naive weight $\P_\nast$ provides a direct estimate of the $x$-distribution of events of the $\nast^{\rm th}$ species. Plots obtained that way are referred to as $\inPlots$: they provide a correct means to reconstruct $\Mtrue_\nast(x)$ only insofar as the variable considered is {\bf in} the set of discriminating variables~$y$. These $\inPlots$ suffer from a major drawback: $x$ being correlated to $y$, the PDFs of~$x$ enter implicitly in the definition of the naive weight, and as a result, the $\tilde\M_\nast$ distributions cannot be used easily to assess the quality of the fit, because these distributions are biased in a way difficult to grasp, 
when the PDFs $\f_i(y)$ are not accurate. For example, let us consider a situation where, in the data sample, 
some events from the $\nast^{\rm th}$ species show up far in the tail 
of the $\M_\nast(x)$ distribution which is implicitly used in the fit. The presence of such events implies that the true distribution $\Mtrue_\nast(x)$ must exhibit a tail 
which is not accounted for by $\M_\nast(x)$. These events would enter in the reconstructed $\inPlot$ $\tilde\M_\nast$ with a very small weight, and they would thus escape detection by the above procedure: $\tilde\M_\nast$ would be close to $\M_\nast$, the distribution assumed for~$x$. Only a mismatch in the core of the $x$-distribution can be revealed with $\inPlots$. Stated differently, the error bars which can be attached to each individual bin of $\tilde\M_\nast$ cannot account for the systematical bias inherent to the $\inPlots$.

%However, one should not disregard the interest of $\inPlots$. For example, one can assess the goodness of fit from the $\inPlots$ of the variable measuring the contribution to $\eLik$ of the $\nast^{\rm th}$ species, $x=\ln\f_{\nast^{\rm th}}(y)$.

\section{The $\sPlot$ technique} 
\label{sec:sPlot}
It was shown in the previous Section that if the 'control' variable~$x$ belongs to the set~$y$ of discriminating variables, one can reconstruct the expected distribution of~$x$ with $\inPlots$. However, the $\inPlots$ are not easy to decipher because knowledge of the~$x$ distribution enters in their construction. 

In this Section is considered the more interesting case where the variable~$x$ is truly a control variable, i.e. where~$x$ does not belong to~$y$. More precisely, the two sets of variables~$x$ and~$y$ are assumed to be uncorrelated: hence, the total PDFs~$\f_i(x,y)$ all factorize into products $\Mtrue_i(x)\f_i(y)$.
 
\subsection{The $\sPlot$ formalism}
\label{sec:sPlotformalism}
One may still consider the above distribution $\tilde\M_\nast$, but this time the naive weight is no longer satisfactory: as shown below, Eq.~(\ref{eq:naiveweightworks}) does not hold. This is because, when summing over the events, the $x$-PDFs~$\Mtrue_j(x)$ appear now on the right hand side of Eq.~(\ref{eq:sumGOTOintegral}), while they are absent in the Likelihood function. However, a simple redefinition of the weights allows to overcome this difficulty. 

Considering the naive weight of Eq.~(\ref{eq:weightxiny}):
\begin{eqnarray}
\left< 
N_\nast\tilde\M_\nast(\bar x)
\right>
&=&\int\int\d y\d x\sum_{j=1}^{\Ns}N_j\Mtrue_j(x)\f_j(y)\delta(x-\bar x)\P_\nast
\nonumber \\
&=&\int\d y\sum_{j=1}^{\Ns}N_j\Mtrue_j(\bar x)\f_j(y)
{N_\nast\f_\nast(y)\over\sum_{k=1}^{\Ns}N_k\f_k(y) }
\nonumber \\
\label{eq:sZut}
&=&N_\nast\sum_{j=1}^{\Ns}\Mtrue_j(\bar x)
\left(N_j\int\d y{\f_\nast(y)\f_j(y)\over\sum_{k=1}^{\Ns}N_k\f_k(y)}\right)
\\
&\neq&N_\nast\Mtrue_\nast(\bar x) ~.
\end{eqnarray}
Indeed, 
as announced, 
the previous procedure does not apply. In effect, the correction term appearing in Eq.~(\ref{eq:sZut})
\begin{equation}
N_j\int\d y{\f_\nast(y)\f_j(y)\over\sum_{k=1}^{\Ns}N_k\f_k(y)}
\end{equation}
is not identical to the kroenecker symbol $\delta_{j\nast}$.
The $\inPlot$ distribution $N_\nast\tilde\M_\nast$ obtained using the naive weight
is a linear combination of the true distributions $\Mtrue_j$.
Only if the~$y$ variable was totally discriminating would one recover the correct answer.
In effect, for a total discrimination, $\f_{j\neq\nast}(y)$ vanishes if $\f_\nast(y)$ is non zero. Thus, the product $\f_\nast(y)\f_j(y)$ is equal to $\f_\nast^2(y)\delta_{j\nast}$, and one gets:
\begin{equation}
N_j\delta_{j\nast}\int\d y{\f_\nast^2(y)\over N_\nast\f_\nast(y)}=\delta_{j\nast} ~.
\end{equation}
But this is purely academic, because, if~$y$ was totally discriminating, the obtention of $\Mtrue_\nast(x)$ would be straightforward: one would just apply cuts on~$y$ to obtain a pure sample of events of the $\nast^{\rm th}$ species and plot them to get $\Mtrue_\nast(x)$.

However, in the case of interest where~$y$ is not totally discriminating, one observes that the correction term is related to the inverse of the covariance matrix, given by the second derivatives of $-\eLik$, which the analysis minimizes:
\begin{equation}
\label{eq:VarianceMatrixDirect}
\V^{-1}_{\nast j}~=~
{\partial^2(-\eLik)\over\partial N_\nast\partial N_j}~=~
\sum_{e=1}^N {\f_\nast(y_e)\f_j(y_e)\over(\sum_{k=1}^{\Ns}N_k\f_k(y_e))^2} ~.
\end{equation}
On average, 
replacing the sum over events by an integral (Eq.~(\ref{eq:sumGOTOintegral})) the variance matrix reads:
\begin{eqnarray}
\label{VarianceMatrixAsymptotic}
\left<
\V^{-1}_{\nast j}
\right>
&=&
\int\int\d y\d x\sum_{l=1}^{\Ns}N_l\Mtrue_l(x)\f_l(y)
{\f_\nast(y)\f_j(y)\over(\sum_{k=1}^{\Ns}N_k\f_k(y))^2}
\nonumber \\
&=&
\int\d y\sum_{l=1}^{\Ns}N_l\f_l(y)
{\f_\nast(y)\f_j(y)\over(\sum_{k=1}^{\Ns}N_k\f_k(y))^2}
\int\d x\Mtrue_l(x)
\nonumber \\
&=&
\label{VarianceMatrix}
\int\d y {\f_\nast(y)\f_j(y)\over\sum_{k=1}^{\Ns}N_k\f_k(y)} ~.
\end{eqnarray}
Therefore, Eq.~(\ref{eq:sZut}) can be rewritten:
\begin{equation}
\label{eq:resZut}
\left<
\tilde\M_\nast(\bar x)
\right>
=
\sum_{j=1}^{\Ns}\Mtrue_j(\bar x)
N_j
\left<
\V^{-1}_{\nast j}
\right>
 ~.
\end{equation}
Inverting this matrix equation, one recovers the distribution of interest:
\begin{equation}
N_\nast \Mtrue_\nast(\bar x)=\sum_{j=1}^{\Ns} 
\left<
\V_{\nast j}
\right>
\left<
\tilde\M_j(\bar x)
\right>
 ~.
\end{equation}
Hence, if the control variable~$x$ is uncorrelated with the discriminating variable, the true distribution of~$x$ can still be reconstructed using the naive weight of Eq.~(\ref{eq:weightxiny}), through a linear combination of the $\inPlots$. This result is better restated as follows. When~$x$ does not belong to the set~$y$, the appropriate weight is not given by Eq.~(\ref{eq:weightxiny}), but is the covariance-weighted quantity (thereafter called sWeight) defined by:
\begin{equation}
\begin{Large}
\label{eq:weightxnotiny}
\fbox{$
\sP_\nast(y_e)={\sum_{j=1}^{\Ns} \V_{\nast j}\f_j(y_e)\over\sum_{k=1}^{\Ns}N_k\f_k(y_e) } $}
\end{Large} ~.
\end{equation}
With this sWeight, the distribution of the control variable~$x$ can be obtained from the $\sPlot$ histogram:
\begin{equation}
\label{eq:masterequation}
{N_\nast}\ _s\tilde\M_\nast(\bar x)\Deltax ~\equiv~ \sum_{\chiant}  \sP_\nast(y_e) ~,
\end{equation}
which reproduces, on average, the true distribution:
\begin{equation}
\label{eq:sPlotsFormula}
\left<
{N_\nast}\ _s\tilde\M_\nast(x)
\right>
~=~ N_\nast \Mtrue_\nast(x) ~.
\end{equation}
If the control variable $x$ exhibits significant correlation with the discriminating variable $y$, the $\sPlots$ obtained with Eq.~(\ref{eq:masterequation}) cannot be compared directly with the pure distributions of the various species. In that case, one must proceed to a Monte-Carlo simulation of the procedure to obtain the expected distributions to which the $\sPlots$ should be compared with.

The fact that the matrix $\V_{ij}$ enters in the definition of the sWeights is enlightening,
and, as discussed in the next Section, this confers nice properties to the $\sPlots$. But this is not the key point. The key point is that Eq.~(\ref{eq:sZut}) is a matrix equation which can be inverted using a numerical evaluation of the matrix based only on data, thanks to Eq.~(\ref{eq:VarianceMatrixDirect}). Rather than computing the matrix by this direct sum over the events, on can use the covariance matrix resulting from the fit, but this option is numerically less accurate than the direct computation\footnote{ Furthermore, when parameters are fitted together with the yields $N_j$, in order to get the correct matrix, one should take care to perform a second fit, where these parameters are frozen.}.

\subsection{$\sPlot$ Properties}
\label{sec:sPlotprop}
Beside satisfying, on the average, the essential asymptotic property Eq.~(\ref{eq:sPlotsFormula}), $\sPlots$ bear properties which hold even under non-asymptotic conditions.

\subsubsection{Normalization}
The distribution $_s\tilde\M_\nast$ 
defined by Eq.~(\ref{eq:masterequation}) is guaranteed to be normalized to unity
and the sum over the species of the $\sPlots$ reproduces the data sample distribution of the control variable.
These two properties are not obvious because,
from expression Eq.~(\ref{eq:weightxnotiny}), neither is it obvious that the sum over the $x$-bins of $N_\nast\ _s\tilde\M_\nast\Deltax$ is equal to $N_\nast$, nor is it obvious that, in each bin, the sum over all species of the expected numbers of events equates to the number of events actually observed. The demonstration uses the three sum rules below.

\begin{enumerate}
\item{} Maximum Likelihood Sum Rule\\
The Likelihood Eq.~(\ref{eq:eLik}) being extremal for $N_j$, one gets the first sum rule:
\begin{equation}
\label{eq:likequation}
\sum_{e=1}^{N}{\f_j(y_e)\over\sum_{k=1}^{\Ns}N_k\f_k(y_e)}=1~,~ \forall j ~.
\end{equation}
\item{} Variance Matrix Sum Rule\\
From Eq.~(\ref{eq:VarianceMatrixDirect}) and Eq.~(\ref{eq:likequation}) one derives:
\begin{equation}
\label{eq:VarianceSumRule}
\sum_{i=1}^{\Ns} N_i \V_{ij}^{-1}~=~
\sum_{i=1}^{\Ns} N_i
\sum_{e=1}^N {\f_i(y_e)\f_j(y_e)\over(\sum_{k=1}^{\Ns}N_k\f_k(y_e))^2} ~=~
\sum_{e=1}^N {\f_j(y_e)\over\sum_{k=1}^{\Ns}N_k\f_k(y_e)} 
=1 ~.
\end{equation}
\item{} Covariance Matrix Sum Rule\\
Multiplying both sides of Eq.~(\ref{eq:VarianceSumRule}) by $\V_{jl}$ and summing over $j$ one gets the sum rule:
\begin{equation}
\label{eq:CoVarianceSumRule}
\sum_{j=1}^{\Ns}\V_{jl}~=~
\sum_{j=1}^{\Ns}\V_{jl}\sum_{i=1}^{\Ns} N_i \V_{ij}^{-1}~=~
\sum_{i=1}^{\Ns}\left(\sum_{j=1}^{\Ns}\V_{ij}^{-1}\V_{jl}\right)N_i~=~
\sum_{i=1}^{\Ns}\delta_{il}N_i ~=~
N_l~.
\end{equation}
\end{enumerate}
It follows that:
\begin{itemize}
\item{}
Each $x$-distribution is properly normalized (cf. Eq.~(\ref{eq:likequation}) and Eq.~(\ref{eq:CoVarianceSumRule})):
\begin{equation}
\label{eq:NormalizationOK}
\sum_{[\Deltax]}N_\nast\ _s\tilde\M_\nast(x)\Deltax~=~ 
\sum_{e=1}^{N} \sP_\nast(y_e)~=~
\sum_{e=1}^{N}{\sum_{j=1}^{\Ns} \V_{\nast j}\f_j(y_e)\over\sum_{k=1}^{\Ns}N_k\f_k(y_e) }~=~
\sum_{j=1}^{\Ns} \V_{\nast j}~=~
N_\nast ~.
\end{equation}

\item{}
The contributions $\sP_j(y_e)$ add up to the number of events actually observed in each $x$-bin. In effect, for any event (cf. Eq.~(\ref{eq:CoVarianceSumRule})) :
\begin{equation}
\label{eq:numberconservation}
\sum_{l=1}^{\Ns} \sP_l(y_e) ~=~
\sum_{l=1}^{\Ns}
{\sum_{j=1}^{\Ns} \V_{l j}\f_j(y_e)\over\sum_{k=1}^{\Ns}N_k\f_k(y_e) }~=~
{\sum_{j=1}^{\Ns} N_j\f_j(y_e)\over\sum_{k=1}^{\Ns}N_k\f_k(y_e) }~=~
1 ~.
\end{equation}
\end{itemize}
Therefore, an $\sPlot$ provides a consistent representation of how 
all events from the various species are distributed in the control variable~$x$.
The contributions to the data sample distribution in~$x$ from the various species are disentangled according to a fit based on the discriminating variable~$y$, provided~$x$ and~$y$ are uncorrelated. Summing up the $\Ns$ $\sPlots$, one recovers the data sample distribution in~$x$,
and summing up the number of events entering in a $\sPlot$ for a given species,
one recovers the yield of the species, 
as it is provided by the fit.

For instance, if one observes an excess of events for a particular $\nast^{\rm th}$ species, 
in a given $x$-bin, 
this excess is effectively accounted for in the number of event $N_\nast$ resulting from the fit. To remove these events (for whatever reason and by whatever means) implies a corresponding decrease in $N_\nast$. It remains to gauge how significant is an anomaly in the $x$-distribution of the $\nast^{\rm th}$ species. This is the subject of the next Section.

\subsubsection{Statistical uncertainties}
\label{sec:StaUnc}
The statistical uncertainty on ${N_\nast}\  _s\tilde\M_\nast(x) \Deltax$ can be defined in each bin by 
\begin{equation}
\label{eq:ErrorPerBin}
\sigma[N_\nast\  _s\tilde\M_\nast(x) \Deltax]~=~\sqrt{\sum_{\chiant} (\sP_\nast)^2} ~.
\end{equation}
The proof that Eq.~(\ref{eq:ErrorPerBin}) holds asymptotically goes as follows: 
\begin{eqnarray}
\left< \left( \sum_{\chiant} \sP_\nast \right) ^2\right>
- \left<\sum_{\chiant} \sP_\nast \right>^2
&=&
 \left< \Nx\right> \left< \sP_\nast^2\right>
+\left< \Nx \left(\Nx-1 \right)\right> \left< \sP_\nast\right>^2
-\left< \Nx\right>^2\left< \sP_\nast\right>^2
\nonumber \\
&=&
 \left< \Nx\right> \left< \sP_\nast^2\right>
+\left( \left< \Nx^2\right>-\left< \Nx\right> \right) \left< \sP_\nast\right>^2
-\left< \Nx\right>^2\left< \sP_\nast\right>^2
\nonumber \\
&=&
 \left< \Nx\right> \left< \sP_\nast^2\right> \nonumber \\
&{\phantom =}&
+\left( \left< \Nx\right> +\left< \Nx\right>^2-\left< \Nx\right> \right) \left< \sP_\nast\right>^2
-\left< \Nx\right>^2\left< \sP_\nast\right>^2
\nonumber \\
&=&
 \left< \Nx\right> \left< \sP_\nast^2\right>
=\left< \sum_{\chiant} \left(\sP_\nast \right)^2 \right>
=\left< \sigma^2[N_\nast\  _s\tilde\M_\nast \Deltax] \right> ~.
\end{eqnarray}
The above asymptotic property is completed by the fact that the sum in quadrature of the uncertainties Eq.~(\ref{eq:ErrorPerBin}) reproduces the statistical uncertainty on the yield $N_\nast$, as it is provided by the fit: $\sigma[N_\nast]\equiv\sqrt{\V_{\nast\nast}}$. The sum over the $x$-bins reads:
\begin{eqnarray}
\sum_{[\Deltax]}\sigma^2[N_\nast\  _s\tilde\M_\nast \Deltax]
&=& \sum_{[\Deltax]}\sum_{\chiant} (\sP_\nast)^2 = \sum_{e=1}^{N}
\left({\sum_{j=1}^{\Ns} \V_{\nast j}\f_j(y_e)\over\sum_{k=1}^{\Ns}N_k\f_k(y_e) }\right)^2
\nonumber \\
&=&
\sum_{j=1}^{\Ns} \sum_{l=1}^{\Ns}
\V_{\nast l}\V_{\nast j}\sum_{e=1}^{N}
{\f_l(y_e)\f_j(y_e)\over(\sum_{k=1}^{\Ns}N_k\f_k(y_e))^2}
\nonumber \\
&=&
\sum_{j=1}^{\Ns} \sum_{l=1}^{\Ns}
\V_{\nast l}\V_{\nast j}\V_{lj}^{-1}
~=~\sum_{l=1}^{\Ns}\V_{\nast l}\delta_{\nast l}
\nonumber \\
\label{eq:SumOfErrors}
&=&
\V_{\nast \nast} ~,
\end{eqnarray}
and more generally, the whole covariance matrix is reproduced:
\begin{equation}
\label{eq:SumOfErrorsij}
\sum_{e=1}^N (\sP_i)(\sP_j)=\V_{ij}~.
\end{equation}
Therefore, for the expected number of events per $x$-bin indicated by the $\sPlots$, 
the statistical uncertainties are straightforward to compute using Eq.~(\ref{eq:ErrorPerBin}).
The later expression is asymptotically correct, and it provides a consistent representation of how 
the overall uncertainty on $N_\nast$ is distributed in~$x$ among the events of the $\nast^{\rm th}$ species.
Because of Eq.~(\ref{eq:SumOfErrorsij}), and since the determination of the yields is optimal when obtained using a Likelihood fit, one can conclude that the $\sPlot$ technique is itself an optimal method to reconstruct distributions of control variables.~\footnote{
This is not the case for $\inPlots$ for which one gets:
\begin{equation}
\sum_{e=1}^N (\P_i)(\P_j)=N_i N_j \V_{ij}^{-1}~.
\end{equation}
Hence, using the fact, that contrary to sWeights, the $\inPlot$ weights of Eq.~(\ref{eq:weightxiny}) are positive definite, one gets: 
\begin{equation}
\sum_{e=1}^N (\P_i)^2
\le
\sum_{e=1}^N (\P_i)(\sum_{j=1}^{\Ns} \P_j)
=
N_i\sum_{j=1}^{\Ns} N_j \V_{ij}^{-1}=N_i \le \V_{ii} ~.
\end{equation}
That is to say that the statistical uncertainties attached to the $\inPlots$ are always not only smaller than the ones resulting from the fit, but even smaller than the statistical uncertainties obtained in a backgound free situation.
}

\subsubsection{Merging $\sPlots$}
As a result of the above, 
two species $i$ and $j$ can be merged into a single species $(i+j)$ 
without having to repeat the fit and recompute the sWeights. The $\sPlot$ of the merged species is just the sum of the two $\sPlots$ obtained by adding the sWeights on an event-by-event basis:
\begin{equation}
N_{(i+j)} \tilde\M_{(i+j)} \Deltax =\sum_{\chiant} (\sP_i+\sP_j) ~.
\end{equation}
The resulting $\sPlot$ has the proper normalization and the proper error bars (Eqs.~(\ref{eq:NormalizationOK}) and~(\ref{eq:SumOfErrorsij})):
\begin{eqnarray}
N_{(i+j)}&=&\sum_{e=1}^N (\sP_i+\sP_j) = N_i+N_j
\\
\sigma^2[N_{(i+j)}]&=&\sum_{e=1}^N (\sP_i+\sP_j)^2 \nonumber \\
&=&\V_{ii}+\V_{jj}+2 \V_{ij} ~=~ \V_{(i+j)(i+j)} ~.
\end{eqnarray}

\subsection{$\sPlot$ implementation}
\label{sec:Cuisine}
This Section is meant to show that using $\sPlot$ is indeed easy.
The different steps to implement the technique are the following:
\begin{enumerate}
\item One is dealing with a data sample in which several species of events are present.
\item A maximum Likelihood fit is performed to obtain the yields $N_i$ of the various species. 
The fit relies on a discriminating variable~$y$ uncorrelated with a control variable~$x$:
the later is therefore totally absent from the fit. 
\item The sWeights $\sP$ are calculated using Eq.~(\ref{eq:weightxnotiny}) where the covariance matrix is obtained by inverting the matrix given by Eq.~(\ref{eq:VarianceMatrixDirect}).
\item Histograms of~$x$ are filled by weighting the events with the sWeights $\sP$. 
The sum of the entries are equal to the yields $N_i$ provided by the fit.
\item Error bars per bin are given by Eq.~(\ref{eq:ErrorPerBin}). 
The sum of the error bars squared are equal to the uncertainties squared $\V_{ii}$ provided by the fit.   
\item The $\sPlots$ reproduce the true distributions of the species in the control variable~$x$, within the above defined statistical uncertainties.
\end{enumerate}
The $\sPlot$ method has been implemented in the ROOT framework under the class TSPlot~\cite{bib:ROOT}.

\subsection{Illustrations}
\label{sec:Illustrations}
\begin{figure}
\begin{center}
  \mbox{{\psfig{file=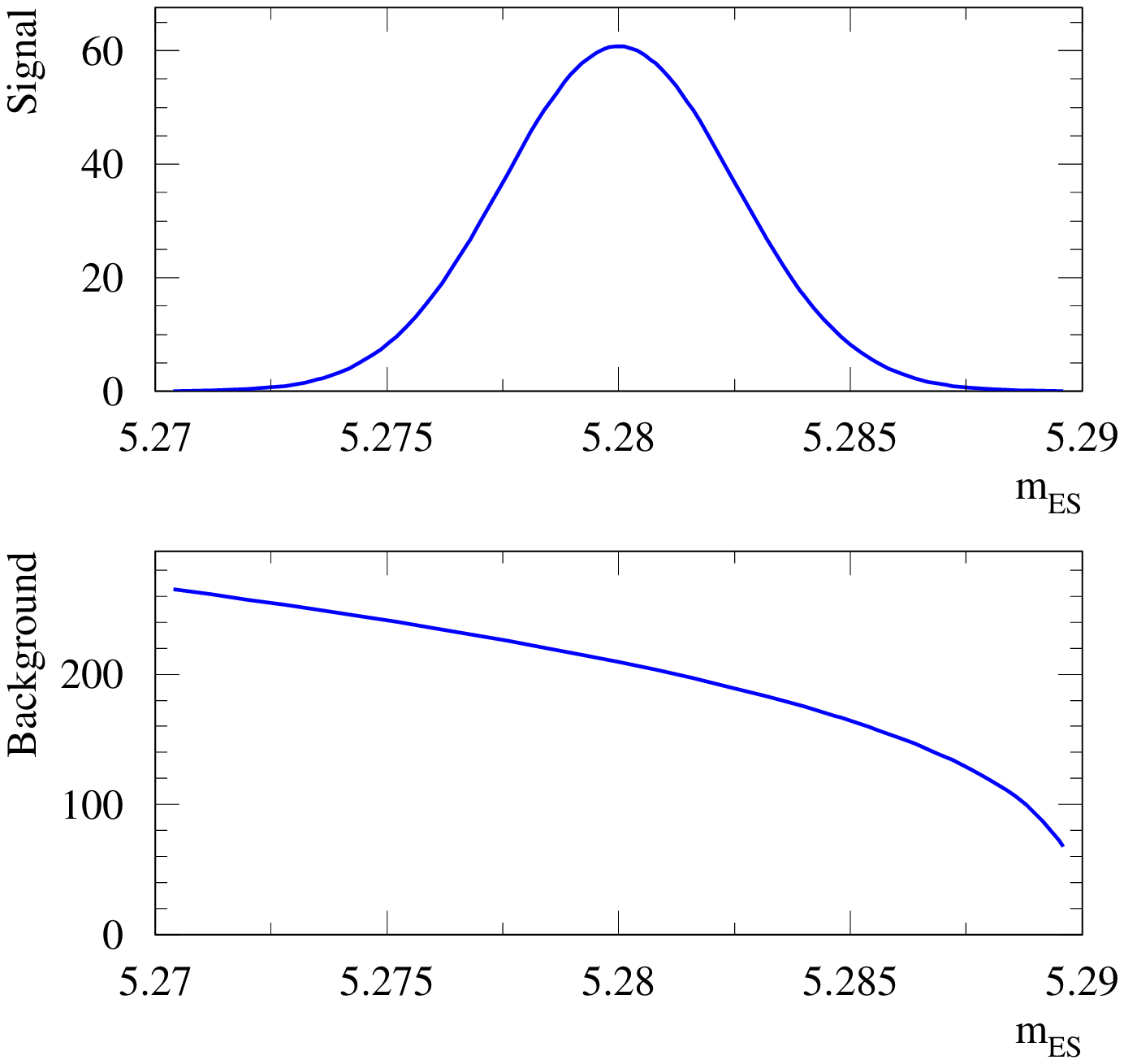,width=0.33\linewidth}}
    {\psfig{file=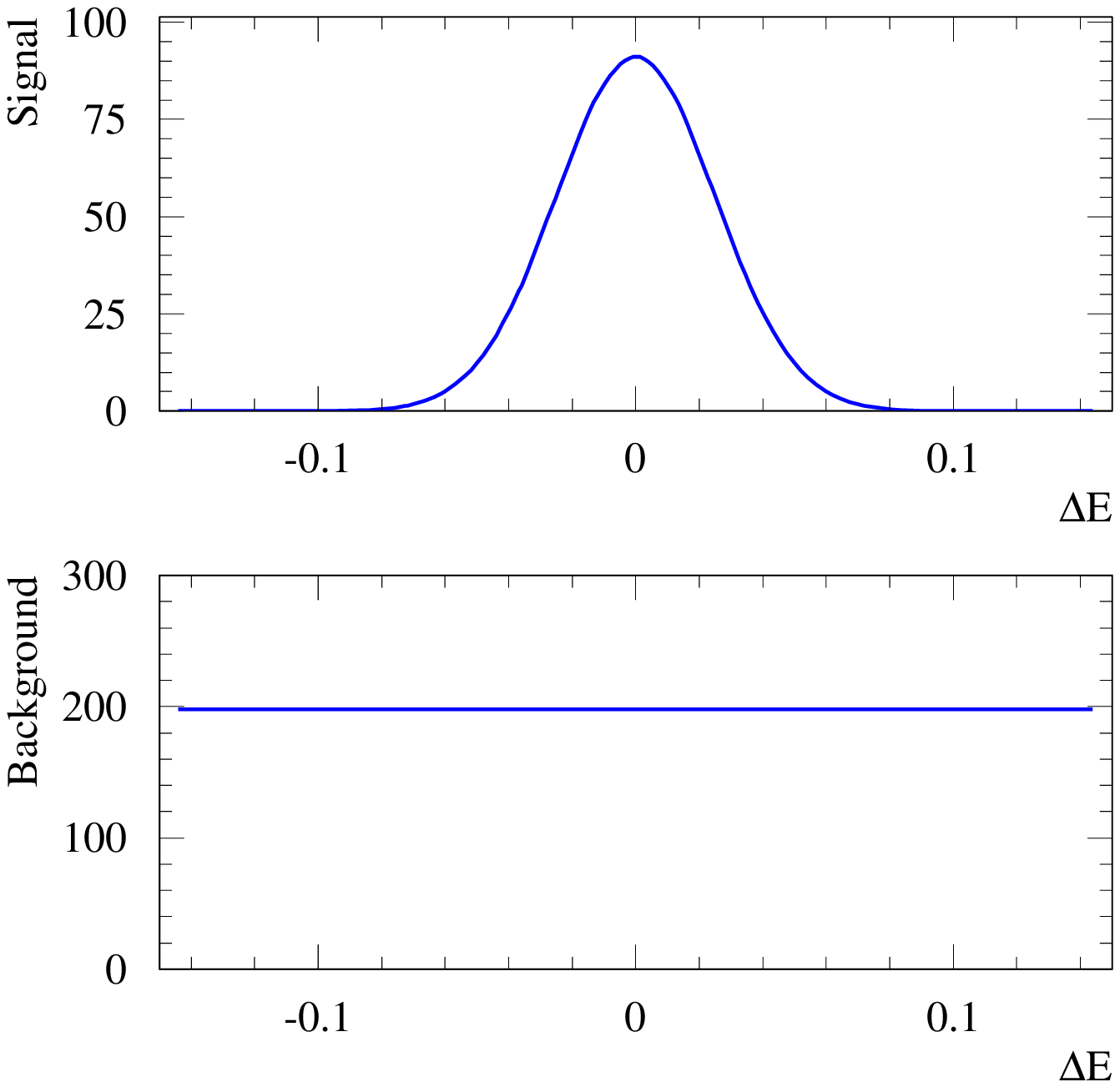,width=0.33\linewidth}}
    {\psfig{file=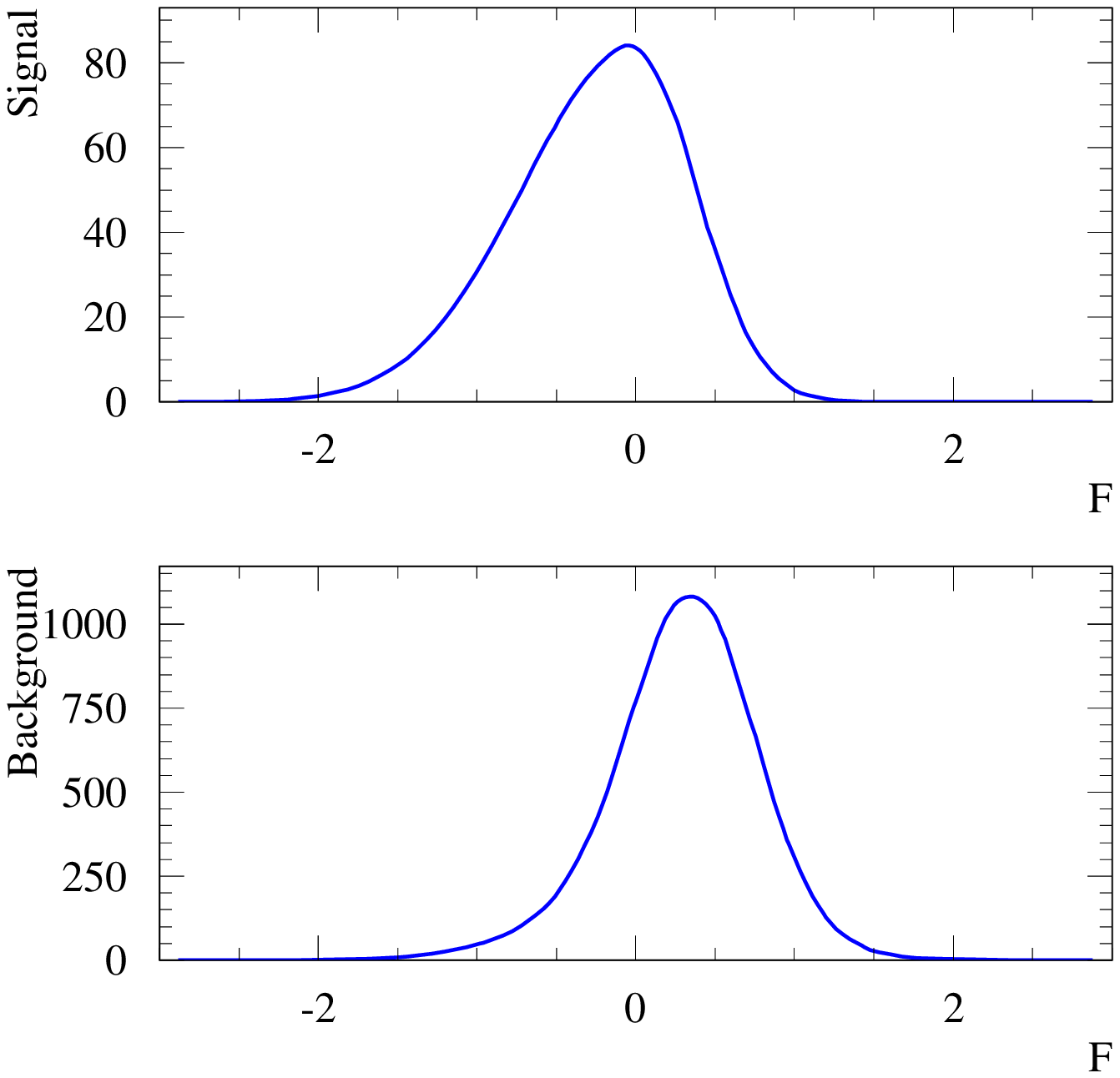,width=0.33\linewidth}}}
\caption{Distributions of the three different discriminating variables available to perform the Likelihood fit: 
$\mes$, $\deltaE$, $\Fisher$. Among the three variables, two are used to perform the fit while one is kept out of the fit to serve the purpose of a control variable. The three distributions on the top (resp. bottom) of the figure correspond to the signal (resp. background). The unit of the vertical axis is chosen such that it indicates the number of entries per bin, if one slices the histograms in 25 bins.}
\label{fig:pdfs}
\end{center}
\end{figure}

To illustrate the technique, one considers in this Section an example derived from the analysis where $\sPlots$ have been first used~\cite{bib:TheseMu} and ~\cite{bib:Kpi} (but see also~\cite{bib:publis}). One is dealing with a data sample in which two species are present: the first is termed signal and the second background. A maximum Likelihood fit is performed to obtain the two yields $N_1$ and $N_2$. The fit relies on two discriminating variables collectively denoted~$y$ which are chosen within three possible variables denoted (following the notations of~\cite{bib:TheseMu}) $\mes$, $\deltaE$ and $\Fisher$.
The variable which is not incorporated in~$y$ is used as a control variable~$x$. The six distributions of the three variables are assumed to be the ones depicted in Fig.~\ref{fig:pdfs}.

\begin{figure}
\begin{center}
  \mbox{{\psfig{file=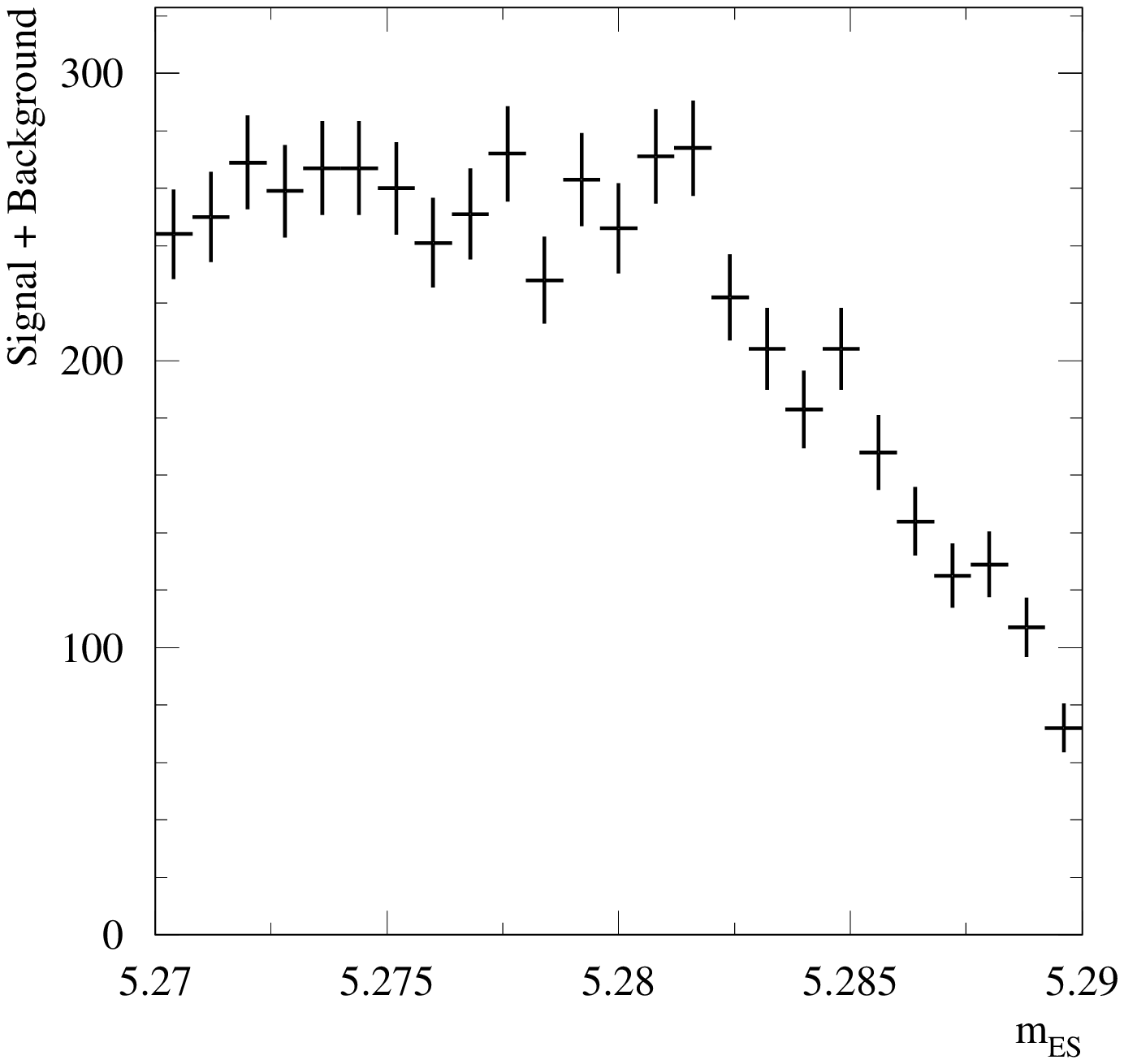,width=0.33\linewidth}}
    {\psfig{file=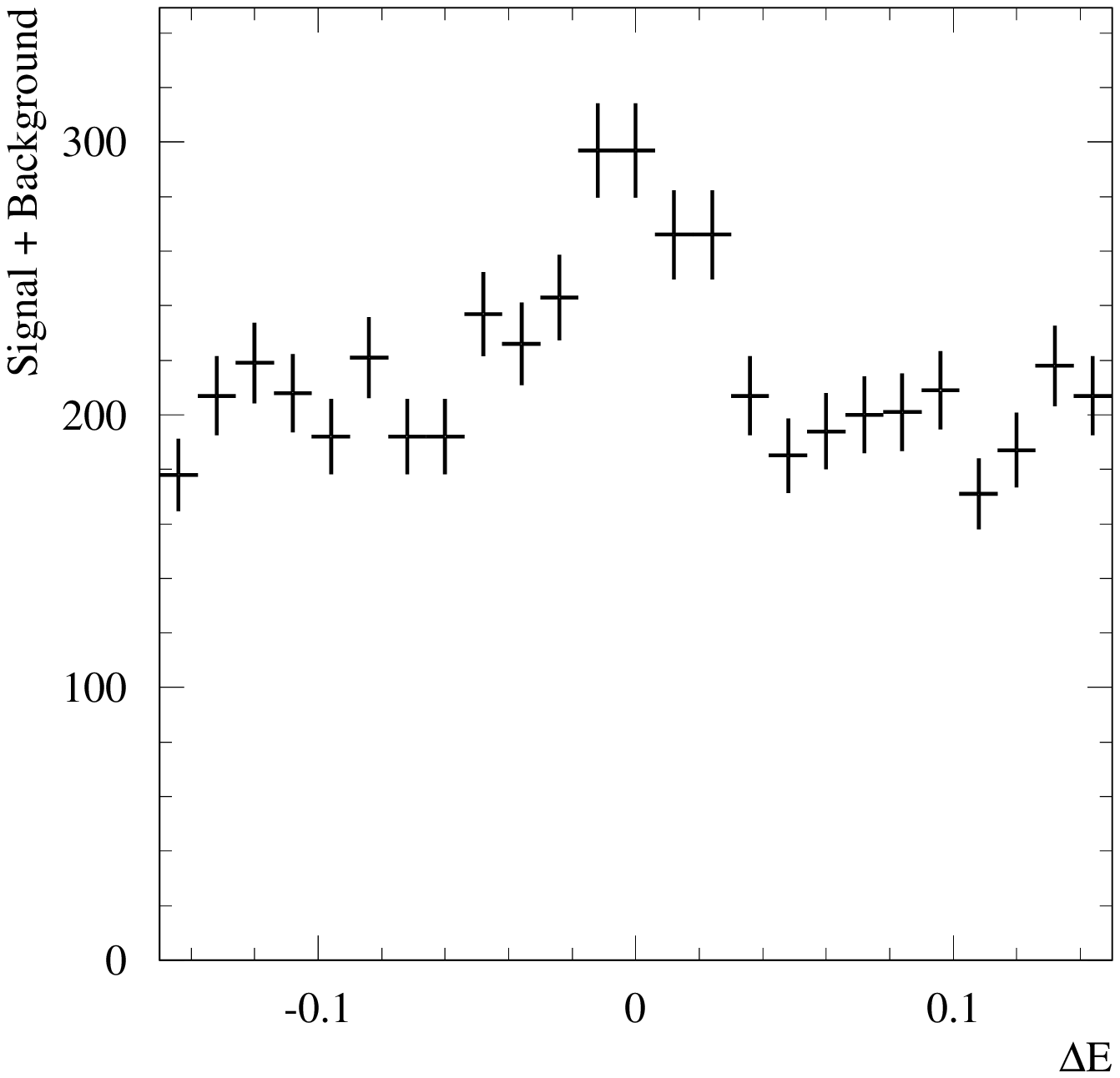,width=0.33\linewidth}}
    {\psfig{file=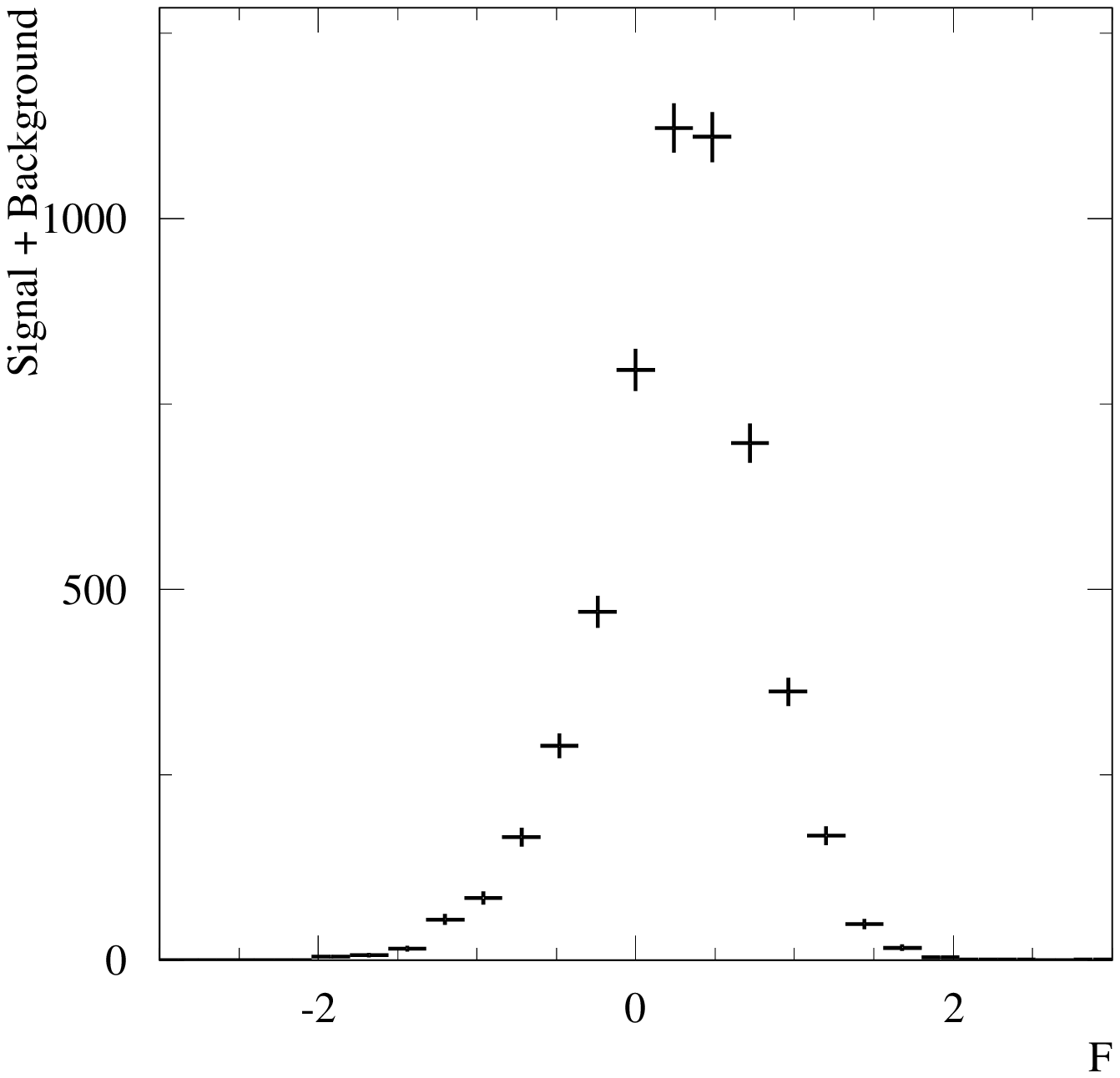,width=0.33\linewidth}}}
\caption{Distributions of the three discriminating variables for signal plus background. The three distributions are the one obtained from a data sample obtained through a Monte Carlo simulation based on the distributions shown in Fig.~\ref{fig:pdfs}.  The data sample consists of 500 signal events and 5000 background events.}
\label{fig:pdfstot}
\end{center}
\end{figure}

A data sample being built through a Monte Carlo simulation based on the distributions shown in Fig.~\ref{fig:pdfs}, one obtains the three distributions of Fig.~\ref{fig:pdfstot}. Whereas the distribution of~$\deltaE$ clearly indicates the presence of the signal, the distribution of $\mes$ and $\Fisher$ are less obviously populated by signal.

\begin{figure}
\begin{center}
  \mbox{\psfig{file=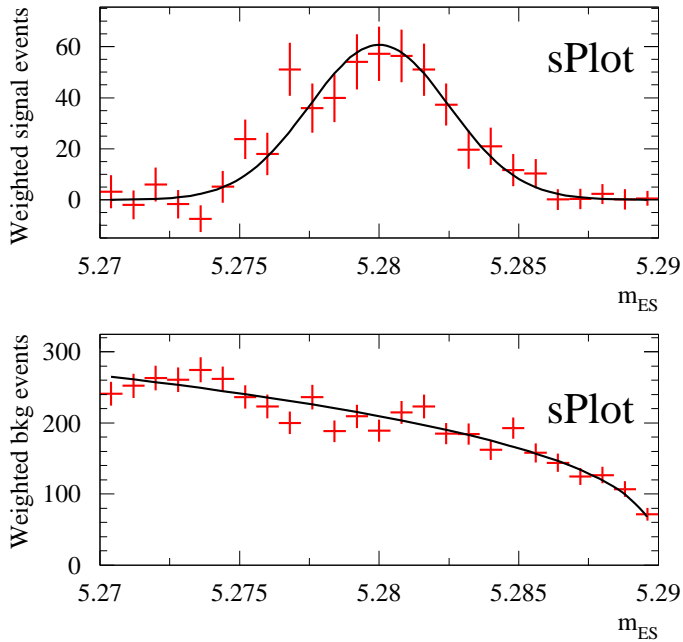,width=0.6\linewidth}}
  \caption{The $\sPlots$ (signal on top, background on bottom) obtained for $\mes$ are represented as dots with error bars. They are obtained from a fit using only information from $\deltaE$ and $\Fisher$. The black curves are the PDFs of $\mes$ of Fig.~\ref{fig:pdfs}: these PDFs are unknown to the fit.}% The histogram shows the distributions of signal and background events actually generated in this Monte Carlo simulation.}
\label{fig:messPlots}
\end{center}
\end{figure}

Chosing $\deltaE$ and $\Fisher$ as discriminating variables to determine $N_1$ and $N_2$ through a maximum Likelihood fit, one builds, for the control variable $\mes$ which is unknown to the fit, the two $\sPlots$ for signal and background shown in Fig.~\ref{fig:messPlots}. For comparison, the PDFs of $\mes$ taken from Fig.~\ref{fig:pdfs} are superimposed on the $\sPlots$. One observes that the $\sPlot$ for signal reproduces correctly the PDF even where the latter vanishes, although the error bars remain sizeable. This results from the almost complete cancellation between positive and negative sWeights: the sum of sWeights is close to zero in the tails while the sum of sWeights squared is not. The occurence of negative sWeights is provided through the appearance of the covariance matrix, and its negative components, in the definition of Eq.~(\ref{eq:weightxnotiny}).

%The histograms superimposed on the $\sPlots$ show the distribution in~$x$ of the events actually generated for the two species. One observes, as indeed they should, that the $\sPlots$ tend to reproduce the histograms, rather than the PDFs. For example, for the signal, in the ninth bin from the left the $\sPlot$ value is too large with respect to the PDF, because of a genuine large statistical fluctuation in this bin, for the Monte Carlo simulated signal: this statistical fluctuation is reproduced by the $sPlot$. 

A word of caution is in order with respect to the error bars. Whereas their sum in quadrature is identical to the statistical uncertainties of the yields determined by the fit, and if, in addition, they are asymptotically correct (cf. Section~\ref{sec:StaUnc}) the error bars should be handled with care for low statistics and/or for too fine binning. This is because the error bars do not incorporate two known properties of the PDFs: PDFs are positive definite and can be non-zero in a given x-bin, even if in the particular data sample at hand, no event is observed in this bin. The latter limitation is not specific to $\sPlots$, rather it is always present when one is willing to infer the PDF at the origin of an histogram, when, for some bins, the number of entries does not guaranty the applicability of the Gaussian regime. In such situations, a satisfactory practice is to attach allowed ranges to the histogram to indicate the upper and lower limits of the PDF value which are consistent with the actual observation, at a given confidence level. Although this is straightforward to implement, even when dealing with sWeighted events, for the sake of simplicity, this subject is not discussed further in the paper.

%Chosing $\mes$ and $\Fisher$ as discriminating variables to determine $N_1$ and $N_2$ through a maximum likelihood fit,
%one builds,
%for the control variable $\deltaE$ which is unknown to the fit, 
%the two $\sPlots$ for signal and background shown in Fig.~\ref{fig:desPlots}.
%For comparison, 
%the PDFs of $\deltaE$ taken from Fig.~\ref{fig:pdfs} are superimposed on the $\sPlots$. 
%
%\begin{figure}
%\begin{center}
%  \mbox{\psfig{file=sdENIM.eps,width=0.6\linewidth}}
%  \caption{The $\sPlots$ (signal on top, background on bottom) obtained for $\deltaE$ are 
%represented as dots with error bars. 
%They are obtained from a fit using only information from $\mes$ and $\Fisher$.
%The black curves are the PDFs of $\deltaE$ of Fig.~\ref{fig:pdfs}:
%these PDFs are unknown to the fit.
%The histogram shows the distributions of signal and background events actually generated in this Monte Carlo simulation.}
%\label{fig:desPlots}
%\end{center}
%\end{figure}

\begin{figure}
\begin{center}
  \mbox{\psfig{file=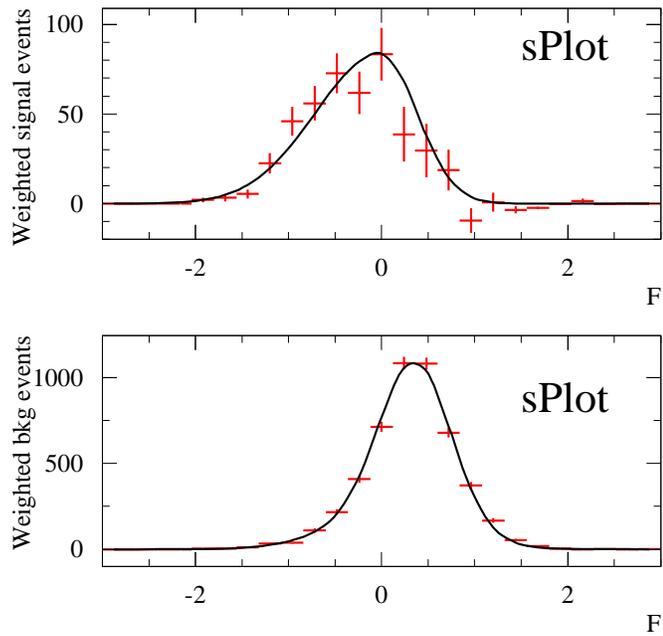,width=0.6\linewidth}}
  \caption{The $\sPlots$ (signal on top, background on bottom) obtained for $\Fisher$ are represented as dots with error bars. They are obtained from a fit using only information from $\mes$ and $\deltaE$. The black curves are the PDFs of $\Fisher$ of Fig.~\ref{fig:pdfs}: these PDFs are unknown to the fit.}% The histogram shows the distributions of signal and background events actually generated in this Monte Carlo simulation.}
\label{fig:FisPlots}
\end{center}
\end{figure}

Chosing $\mes$ and $\deltaE$ as discriminating variables to determine $N_1$ and $N_2$ through a maximum Likelihood fit, one builds, for the control variable $\Fisher$ which is unknown to the fit, the two $\sPlots$ for signal and background shown in Fig.~\ref{fig:FisPlots}. For comparison, the PDFs of $\Fisher$ taken from Fig.~\ref{fig:pdfs} are superimposed on the $\sPlots$. In the $\sPlot$ for signal one observes that error bars are the largest in the~$x$ regions where the background is the largest.

\subsection{Application: efficiency corrected yields}
\label{sec:DBR}
Beside providing a convenient and optimal tool to cross-check the analysis by allowing distributions
of control variables 
to be reconstructed and then compared with expectations, 
the $\sPlot$ formalism can be applied also to extract physics results, which would otherwise be difficult to obtain. For example, one may be willing to explore some unknown physics involved in the distribution of a variable $x$. Or, one may be interested to correct a particular yield provided by the Likelihood fit from a selection efficiency which is known to depend on a variable $x$, for which the PDF is unknown.

To be specific, 
one can take the example of a three body decay analysis of a species, the signal, 
polluted by background, while the signal PDF inside the two-dimensional Dalitz plot is not known, because of unknown contributions of resonances, continuum and an interference pattern. Since the $x$-dependence of the selection efficiency $\epsilon(x)$ can be computed without {\it a priori} knowledge of the $x$-distributions, one can build the efficiency corrected two-dimensional $\sPlots$ (cf. Eq.~(\ref{eq:masterequation})):
\begin{equation}
\label{eq:Dalitzcorrected}
{1\over\epsilon(\bar{x})} {N_\nast}~_s\tilde\M_\nast(\bar{x}) \delta x = \sum_{\chiant} {1\over\epsilon(x_e)} \sP_\nast(y_e) ~,
\end{equation}
and compute the efficiency corrected yields:
\begin{equation}
\label{eq:br}
N_\nast^\epsilon= \sum_{e=1}^N {\sP_\nast(y_e)\over\epsilon(x_e)} ~.
\end{equation}
Analyses can then use the $\sPlot$ formalism for validation purposes, 
but also, using Eq.~(\ref{eq:Dalitzcorrected}) and Eq.~(\ref{eq:br}), 
to probe for resonance structures and to measure branching ratios.

\section{Conclusion}
The technique presented in this paper applies 
when one examines a data sample originating from different sources of events:
using a set $y$ of discriminating variables,
a Likelihood fit is performed on the data sample to determine the yields of the sources. 
By building $\sPlots$, 
one can reconstruct the distributions of variables, separately for each source present in the data sample,
provided the variables are uncorrelated with the set~$y$ used in the fit. 
Although no cut is applied 
(hence, the $\sPlot$ of a given species represents the whole statistics of this species) 
the distributions obtained are pure (free from the potential background arising from the other species) 
in a statistical sense. 
The more discriminating the discriminating variables~$y$, the clearer the $\sPlot$ is. 
The technique is straightforward to implement and features several nice properties: 
both the normalizations and the statistical uncertainties of the $\sPlots$ reflect the fit ouputs.

\newpage
\appendix
\section{Pedagogical examples}
\label{sec:pedagogicalexamples}
The purpose of this Appendix is to detail in simplified situations how and why $\sPlot$ works. One begins with the simplest situation and proceed to more complex ones.

\subsection{Simple cut-and-count analysis}
\label{sec:CutandCountAnalysis}
In this Section, a very simple situation is considered where the proper way to reconstruct signal and background distributions for a control variable~$x$ is obvious from the start. The purpose is to observe the $\sPlot$ technique at work, when one knows beforehand what the outcome should be. 

One considers a data sample consisting of $\Ns=2$ species: species 1 is referred to as the signal and species 2 as the background. A unique discriminating variable $y\in[0,1]$ is used in the fit. One further assumes that:
\begin{itemize}
\item{}
the signal distribution is the step-function: 
\begin{eqnarray}
\f_1(y<y_0)&=&0
\\
\f_1(y\ge y_0)&=&(1-y_0)^{-1}~,
\end{eqnarray}
\item{} 
the background distribution is uniform in the full range: 
\begin{equation}
\f_2(y)=1 ~.
\end{equation}
\end{itemize}
Therefore, one is dealing with a cut-and-count analysis: there is a pure background side-band for $y<y_0$, and the shapes of the signal and background distributions offer no discriminating power in the region where the signal is present, for $y\ge y_0$. Denoting $N$ the total number of events present in the data sample, $N_<$ the number of events located below $y_0$, and $N_>$ the number of events located above $y_0$: 
\begin{enumerate}
\item{}
the expected number of background and signal events can be deduced without any fit, from the sideband:
\begin{eqnarray}
\label{eq:N2number}
N_2&=&{1\over y_0} N_<
\\
\label{eq:N1number}
N_1&=&-{1-y_0\over y_0} N_< + N_>~,
\end{eqnarray}
\item{}
$N_<$ and $N_>$ being two independent numbers of events, the covariance matrix can be deduced directly from Eqs.~(\ref{eq:N2number})-(\ref{eq:N1number}):
\begin{equation}
\label{eq:covariance}
\V= \left(
                \begin{array}{cc}
        \displaystyle{N_>+\left({1-y_0\over y_0}\right)^2 N_<}
                & \quad
        \displaystyle{-{1-y_0\over y_0^2} N_< }
                \\ \\
        \displaystyle{-{1-y_0\over y_0^2} N_<} 
                & \quad
        \displaystyle{{1\over y_0^2}N_<}
                \end{array}
    \right)~,
\end{equation}
\item{}
denoting $\delta N_<^x$ the number of events in a given $x$-bin, with $y \le y_0$, the background distribution $\M_2(x)$ can also be deduced by a mere rescaling of $\delta N_<^x$, as in Eq.~(\ref{eq:N2number}):
\begin{equation}
\label{eq:back}
N_2\ \M_2(x)\Deltax={\delta N_<^x\over y_0} ~.
\end{equation}
Similarly to Eq.~(\ref{eq:N1number}), the signal distribution is given by:
\begin{equation}
\label{eq:sig}
N_1\ \M_1(x)\Deltax=-(1-y_0)N_2\ \M_2(x)+\delta N_>^x ~,
\end{equation}
that is to say, one can obtain the signal distribution from the (mixed) events populating the domain $y\ge y_0$, if one subtracts the contribution of background events, which is known from Eq.~(\ref{eq:back}). Stated differently, one is lead to assign the negative weight $-(1-y_0)/y_0$ to those events in the $x$-bin which satisfy $y\le y_0$.
\end{enumerate}
Whereas in such a simple situation the use of the $\sPlot$ formalism would be awkward, the latter should reproduce the above obvious results, and indeed it does. The proof goes as follows:
\begin{enumerate}
\item{}
denoting $\f_i(0)$ (resp. $\f_i(1)$) the value taken by the PDF of species $i$ for $y \le y_0$ (resp. $y > y_0$), Eq.~(\ref{eq:likequation}) reads:
\begin{eqnarray}
1
&=&\sum_{e=1}^{N}{\f_1(y_e)\over\sum_{k=1}^{\Ns}N_k\f_k(y_e)}
~=~
N_< {\f_1(0)\over N_1 \f_1(0) + N_2 \f_2(0)}
+
N_> {\f_1(1)\over N_1 \f_1(1) + N_2 \f_2(1)}
\nonumber \\
&=&
{N_>(1-y_0)^{-1}\over N_1 (1-y_0)^{-1} + N_2 }
\\
1&=&\sum_{e=1}^{N}{\f_2(y_e)\over\sum_{k=1}^{\Ns}N_k\f_k(y_e)}
~=~
N_< {\f_2(0)\over N_1 \f_1(0) + N_2 \f_2(0)}
+
N_> {\f_2(1)\over N_1 \f_1(1) + N_2 \f_2(1)}
\nonumber \\
&=&
{N_<\over  N_2 }
+
{N_>\over N_1 (1-y_0)^{-1} + N_2 } ~.
\end{eqnarray}
The first equation yields:
\begin{equation}
\label{eq:useafter}
N_1 (1-y_0)^{-1} + N_2 = N_> (1-y_0)^{-1}
\end{equation}
and thus, for the second equation:
\begin{equation}
1={N_<\over  N_2 }
+1-y_0 ~,
\end{equation}
which leads to Eqs.~(\ref{eq:N2number})-(\ref{eq:N1number}).
\item{}
similarly, Eq.~(\ref{eq:VarianceMatrixDirect}) yields
\begin{equation}
\label{variance}
\V^{-1}= \left(
        \begin{array}{cc}
        \displaystyle{1\over N_>}
                & \quad \displaystyle{1-y_0\over N_>} \\ \\
        \displaystyle{1-y_0\over N_>}
                & \quad \displaystyle{(1-y_0)^2\over N_>}+{y_0^2\over N_<}
        \end{array}
        \right) ~.
\end{equation}
For example, using Eq.~(\ref{eq:useafter}), the $\V_{11}$ component is computed as follows:
\begin{equation}
\V^{-1}_{11} ~=~
\sum_{e=1}^{N}{\f_1(y_e)\f_1(y_e)\over(\sum_{k=1}^{\Ns}N_k\f_k(y_e))^2} ~=~
N_>{(1-y_0)^{-2}\over (N_1(1-y_0)^{-1}+N_2)^2} ={1\over N_>} ~.
\end{equation}
And similarly for the other components. Inverting $\V^{-1}$ one gets Eq.~(\ref{eq:covariance}).
\item{} Eq.~(\ref{eq:masterequation}) then reproduces Eqs.~(\ref{eq:back})-(\ref{eq:sig}). Namely:
\begin{eqnarray}
N_1\ \M_1(x)\Deltax
&=&
\sum_{\chiant}
{\V_{11}\f_1(y_e)+\V_{12}\f_2(y_e)\over\sum_{k=1}^{\Ns}N_k\f_k(y_e)}
\nonumber \\
&=&
\delta N_<^x {\V_{11}\f_1(0)+\V_{12}\f_2(0)\over N_1 \f_1(0)+N_2\f_2(0)}
+
\delta N_>^x {\V_{11}\f_1(1)+\V_{12}\f_2(1)\over N_1 \f_1(1)+N_2\f_2(1)}
\nonumber \\
&=&
\delta N_<^x {\V_{12}\over N_2}
+
\delta N_>^x {\V_{11}(1-y_0)^{-1}+\V_{12}\over N_1(1-y_0)^{-1}+N_2}
\nonumber \\
&=&
\delta N_<^x {-{1-y_0\over y_0^2}N_<\over N_< y_0^{-1}}
\nonumber \\
&\phantom{=}&
+
\delta N_>^x {(N_>+({1-y_0\over y_0})^2N_<)(1-y_0)^{-1}-{1-y_0\over y_0^2}N_<\over N_> (1-y_0)^{-1}}
\nonumber \\
&=&-{1-y_0\over y_0}\delta N_<^x+\delta N_>^x
\end{eqnarray}
and:
\begin{eqnarray}
N_2\ \M_2(x)\Deltax
&=&
\sum_{\chiant}
{\V_{21}\f_1(y_e)+\V_{22}\f_2(y_e)\over\sum_{k=1}^{\Ns}N_k\f_k(y_e)}
\nonumber \\
&=&
\delta N_<^x {\V_{21}\f_1(0)+\V_{22}\f_2(0)\over N_1 \f_1(0)+N_2\f_2(0)}
+
\delta N_>^x {\V_{21}\f_1(1)+\V_{22}\f_2(1)\over N_1 \f_1(1)+N_2\f_2(1)}
\nonumber \\
&=&
\delta N_<^x {\V_{22}\over N_2}
+
\delta N_>^x {\V_{21}(1-y_0)^{-1}+\V_{22}\over N_1(1-y_0)^{-1}+N_2}
\nonumber \\
&=&
\delta N_<^x {{1\over y_0^2}N_<\over N_< y_0^{-1}}
\nonumber \\
&\phantom{=}&
+
\delta N_>^x {-{1-y_0\over y_0^2}N_<(1-y_0)^{-1}+{1\over y_0^2}N_<\over N_> (1-y_0)^{-1}}
\nonumber \\
&=&{\delta N_<^x\over y_0} ~.
\end{eqnarray}

\item{}
it can be shown as well that Eqs.~(\ref{eq:VarianceSumRule})-(\ref{eq:CoVarianceSumRule})-(\ref{eq:NormalizationOK})-(\ref{eq:numberconservation})-(\ref{eq:SumOfErrorsij}) hold.

\end{enumerate}
Therefore, in this very simple situation where the problem of reconstructing the distributions of signal and background events is glaringly obvious, the $\sPlot$ formalism reproduces the expected results.

\subsection{Extended cut-and-count analysis}
\label{ExtendedCutandCountAnalysis}
The above example of the previous Section~\ref{sec:CutandCountAnalysis} is a very particular case of a more general situation where the $y$-range is split into $n_y$ slices inside which one disregards the shape of the distributions of the species, whether these distributions are the same or not. Using greek letters to index the $y$-slices, this amounts to replacing the $\f_i(y)$ PDFs by step functions with constant values. For each $y$-bin $\F_i^\alpha$, these constant values are defined by the integral over the $y$-bin $\alpha$:
\begin{eqnarray}
\label{eq:Fi}
\f_i(y)\rightarrow \F_i^\alpha&=&\int_\alpha \f_i(y)\d y\\
\label{FiNorm}
\sum_{\alpha=1}^{n_y}\F_i^\alpha&=&1~.
\end{eqnarray}
With this notation, the number of events $\bar N_\alpha$ expected in the slice $\alpha$ is given by:
\begin{equation}
\label{eq:Nalpha}
\bar N_\alpha=\sum_{i=1}^{\Ns} N_i\F_i^\alpha ~.
\end{equation}
To make particularly obvious what must be the outcome of the $\sPlot$ technique, 
in the previous Section it was assumed that $n_y=\Ns=2$,
and that the signal was utterly absent in one of the two $y$-slices: 
$\F^1_1=0$, $\F^2_1=1$, $\F^1_2=y_0$ and $\F^2_2=1-y_0$.

Below one proceeds in two steps, first considering the more general case where only $n_y=\Ns$ is assumed (Section~\ref{sec:stepone}), then  considering the extended cut-and-count analysis where $n_y>\Ns$ (Section~\ref{sec:steptwo}). Since the general case discussed in the presentation of the $\sPlot$ formalism corresponds to the limit $n_y\rightarrow\infty$, what follows amounts to a step-by-step new derivation of the technique.

\subsubsection{Generalized cut-and-count analysis: $n_y=\Ns$}
\label{sec:stepone}
When the number of $y$-slices equals the number of species, the solution remains obvious, if the $\Ns\times\Ns$ matrix $\F_i^\alpha$ is invertible (if not, the $N_i$ cannot be determined). In that case, one can identify the expected numbers of events $\bar N_\alpha$ with the observed number of events $N_\alpha$, and thus:
\begin{enumerate}
\item{}
one recovers the expected number of events $N_i$ from the numbers of events $N_\alpha$ observed in the $n_y$ slice, by inverting Eq.~(\ref{eq:Nalpha}):
\begin{equation}
\label{eq:Nalphai}
N_i=\sum_{\alpha=1}^{\Ns} N_\alpha(\F^{-1})_i^\alpha ~,
\end{equation}
\item{}
the number $N_\alpha$ being statistically independent,
one obtains directly the covariance matrix:
\begin{equation}
\label{eq:covariancegeneral}
\V_{ij}=\sum_{\alpha=1}^{\Ns} N_\alpha (\F^{-1})_i^\alpha(\F^{-1})_j^\alpha ~,
\end{equation}
\item{}
similarly to Eq.~(\ref{eq:Nalpha}), the number of events $\delta N_\alpha^x$ observed in the $y$-slice $\alpha$ and in the bin~$x$ of width $\delta x$ is given by:
\begin{equation}
\label{Nalphax}
\delta N_\alpha^x=\sum_{i=1}^{\Ns} N_i\Mtrue_i(x)\delta x\F_i^\alpha
\end{equation}
and thus, the $x$-distribution of species $i$ is:
\begin{equation}
\label{eq:Nalphaix}
\delta N_i^x\equiv N_i \Mtrue_i(x)\delta x=\sum_{\alpha=1}^{\Ns} \delta N_\alpha^x(\F^{-1})_i^\alpha ~.
\end{equation}

\end{enumerate}
It remains to be shown that Eq.~(\ref{eq:Nalphaix}) is reproduced using the $\sPlot$ formalism. 
First, using Eq.~(\ref{eq:Fi}) and Eq.~(\ref{eq:Nalpha}), one observes that: 
\begin{eqnarray}
\sum_{e=1}^N{\f_i(y_e)\over\sum_{k=1}^{\Ns} N_k \f_k(y_e)}
&\rightarrow&
\sum_{\alpha=1}^{\Ns} {N_\alpha}
{\F_i^\alpha\over\sum_{k=1}^{\Ns} N_k \F_k^\alpha}
~=~
\sum_{\alpha=1}^{\Ns} N_\alpha 
{\F_i^\alpha\over N_\alpha}
~=~
\sum_{\alpha=1}^{\Ns} \F_i^\alpha
=1 ~,
\end{eqnarray}
which shows that the obvious solution Eq.~(\ref{eq:Nalphai}) is the one which maximizes the
extended log-Likelihood. Similarly:
\begin{equation}
\V_{ij}^{-1}
=\sum_{e=1}^N{\f_i(y_e)\f_j(y_e)\over(\sum_{k=1}^{\Ns} N_k \f_k(y_e))^2}
\rightarrow
\sum_{\alpha=1}^{\Ns} N_\alpha {\F_i^\alpha\F_j^\alpha\over N_\alpha^2}
~=~\sum_{\alpha=1}^{\Ns} {1\over N_\alpha} \F_i^\alpha\F_j^\alpha ~,
\end{equation}
which inverse is given by Eq.~(\ref{eq:covariancegeneral}), and thus:
\begin{equation}
N_i\ _s\tilde\M_i(x)\delta x
\rightarrow
\sum_{\alpha=1}^{\Ns} \delta N_\alpha^x  
{\sum_j \V_{ij}\F_j^\alpha\over N_\alpha}
~=~
\sum_{\alpha=1}^{\Ns} \delta N_\alpha^x(\F^{-1})^\alpha_i ~.
\end{equation} 
The $\sPlot$ formalism reproduces Eq.~(\ref{eq:Nalphaix}).

\subsubsection{Extended cut-and-count analysis: $n_y>\Ns$}
\label{sec:steptwo}
In the more general situation where the number of $y$-slices is larger than the number of species, there is no blatant solution neither for determining the $N_i$, nor for reconstructing the $x$-distribution of each species (in particular, Eq.~(\ref{eq:Nalphai}) is lost). Because of this lack of an obvious solution, what follows is a rephrasing of the derivation of the $\sPlots$, but taking a different point of view, and in the case where the $y$-distributions are binned.

The best determination of the $N_i$ (here as well as in the previous simpler situations) is provided by the Likelihood method which yields (cf. Eq.~(\ref{eq:likequation})):
\begin{equation}
\label{eq:Nibinned}
\sum_{\alpha=1}^{n_y}{N_\alpha\F^\alpha_i\over\sum_{k=1}^{\Ns} N_k \F^\alpha_k}=1
\ \ \ , \forall i
\end{equation}
with a variance matrix (cf. Eq.~(\ref{eq:VarianceMatrixDirect})):
\begin{equation}
\V_{ij}^{-1}=\sum_{\alpha=1}^{n_y}
N_\alpha{\F^\alpha_i\F^\alpha_j\over(\sum_{k=1}^{\Ns} N_k \F^\alpha_k)^2} ~,
\end{equation}
from which one computes the covariance matrix $\V_{ij}$.
Instead of Eq.~(\ref{eq:Nalphai}) the number of events $N_i$ provided by Eq.~(\ref{eq:Nibinned})
is shown below to satisfy the equality (cf. Eq.~(\ref{eq:NormalizationOK})):
\begin{equation}
\label{eq:NiMaster}
N_i=\sum_{\alpha=1}^{n_y} N_\alpha
\ (\sP)_i^\alpha ~,
\end{equation}
where the matrix element $(\sP)_i^\alpha$ is the sWeight (Eq.~(\ref{eq:weightxnotiny})) for species $i$ of events with $y_e$ lying in the $y$-slice $\alpha$, namely:
\begin{equation}
\label{weightxnotinyBinned}
(\sP)_i^\alpha ~=~{\sum_{j=1}^{\Ns} \V_{ij} \F^\alpha_j
\over\sum_{k=1}^{\Ns} N_k\F^\alpha_k} ~.
\end{equation}
The identity of Eq.~(\ref{eq:NiMaster}) is not asymptotic, it holds even for finite statistics, since the contractions with $\V_{li}^{-1}$ of both the left- and right-hand sides yield the same result. Indeed (Eq.~(\ref{eq:VarianceSumRule})):
\begin{equation}
\sum_{i=1}^{\Ns}  N_i\V_{li}^{-1} ~=~
\sum_{i=1}^{\Ns} \sum_{\alpha=1}^{n_y}
{N_\alpha\F^\alpha_l N_i\F^\alpha_i\over(\sum_{k=1}^{\Ns} N_k \F^\alpha_k)^2}
~=~
\sum_{\alpha=1}^{n_y}{N_\alpha\F^\alpha_l(\sum_{i=1}^{\Ns}N_i\F^\alpha_i)\over(\sum_{k=1}^{\Ns} N_k \F^\alpha_k)^2}
~=~
\sum_{\alpha=1}^{n_y}{N_\alpha\F^\alpha_l\over\sum_{k=1}^{\Ns} N_k \F^\alpha_k}~=~1 ~,
\end{equation}
which is identical to:
\begin{equation}
\sum_{i=1}^{\Ns} \V_{li}^{-1} \sum_{\alpha=1}^{n_y} N_\alpha {\sum_{j=1}^{\Ns} \V_{ij} \F^\alpha_j \over\sum_{k=1}^{\Ns} N_k\F^\alpha_k}
~=~ \sum_{\alpha=1}^{n_y} N_\alpha 
{\sum_{j=1}^{\Ns}(\sum_{i=1}^{\Ns} \V_{li}^{-1} \V_{ij}) \F^\alpha_j
\over\sum_{k=1}^{\Ns} N_k\F^\alpha_k} 
~=~
\sum_{\alpha=1}^{n_y} 
{N_\alpha \F^\alpha_l
\over\sum_{k=1}^{\Ns} N_k\F^\alpha_k} 
~=~1 ~.
\end{equation}
Since Eq.~(\ref{eq:NiMaster}) holds for the complete sample of events,
it must hold as well for any sub-sample, 
provided the splitting into sub-samples is not correlated with the
variable~$y$. Namely, for all $x$-bin, one is guaranteed to observe, on average, the same relationship between the numbers of events $\delta N_i^x$ and $\delta N_\alpha^x$. The $\sPlot$ obtained from the weighted sum
\begin{equation}
\label{eq:NiMasterBinx}
\delta N_i^x=\sum_{\alpha=1}^{n_y} \delta N_\alpha^x 
\ (\sP)_i^\alpha ~,
\end{equation} 
is an unbiased estimator of the true distribution of~$x$ for species~$i$.
One can provide a direct proof that the above $\sPlot$ of Eq.~(\ref{eq:NiMasterBinx}) reproduces the true distribution by following the same line which leads to Eq.~(\ref{eq:resZut}). On average, using successively:
\begin{equation}
\left< \delta N_\alpha^x\right> =\sum_{l=1}^{\Ns} N_l\Mtrue_l(x)\F^\alpha_l\delta x
\end{equation}
and hence:
\begin{equation}
\left< N_\alpha \right>
~=~
\sum_x \left< \delta N_\alpha^x \right> ~=~
\sum_x\Mtrue_l(x)\sum_{k=1}^{\Ns} N_k\F^\alpha_k ~=~
\sum_{k=1}^{\Ns} N_k\F^\alpha_k ~,
\end{equation}
one gets:
\begin{eqnarray}
\left<
\sum_{\alpha=1}^{n_y} \delta N_\alpha^x \ (\sP)_i^\alpha
\right>
&=&
\sum_{\alpha=1}^{n_y} 
\left(\sum_{l=1}^{\Ns} N_l\Mtrue_l(x)\F^\alpha_l\delta x\right) 
{\sum_{j=1}^{\Ns} \V_{ij} \F^\alpha_j
\over\sum_{k=1}^{\Ns} N_k\F^\alpha_k}
\nonumber \\
&=&
\delta x\sum_{l=1}^{\Ns} N_l\Mtrue_l(x)
\left(
\sum_{j=1}^{\Ns} \V_{ij}
\sum_{\alpha=1}^{n_y}{\F^\alpha_j\F^\alpha_l \over\sum_{k=1}^{\Ns} N_k\F^\alpha_k}
\right)
\nonumber \\
&=&
\delta x\sum_{l=1}^{\Ns} N_l\Mtrue_l(x)
\left(
\sum_{j=1}^{\Ns} \V_{ij}
\sum_{\alpha=1}^{n_y}N_\alpha{\F^\alpha_j\F^\alpha_l \over(\sum_{k=1}^{\Ns} N_k\F^\alpha_k)^2}
\right)
\nonumber \\
&=&
\delta x\sum_{l=1}^{\Ns} N_l\Mtrue_l(x)
\left(
\sum_{j=1}^{\Ns} \V_{ij} \V^{-1}_{jl}
\right)
\nonumber \\
&=&
N_i\Mtrue_i(x)\delta x
\equiv 
\left<
\delta N_i^x
\right>
~,
\end{eqnarray}
which concludes the discussion of the situation where the $y$-distributions are step functions.

\section{Extended $\sPlots$: a species is known (fixed)}
\label{sectionextendedsplots}
It may happen that the yields of some species are not derived from the data sample at hand, 
but are taken to be known from other sources of information. 
Here, one denotes collectively as species '0' the overall component of such species. The number of expected events for species '0', $N_0$, being assumed to be known, is held fixed in the fit. In this Section, the indices $i,j...$ run over the $\Ns$ species for which the yields $N_i$ are fitted, the fixed species '0' being excepted ($i,j...\neq 0$).

One can meet various instances of such a situation. Two extreme cases are:
\begin{enumerate}
\item{} the species '0' is very well known, such that the information on it contained by the data sample at hand is irrelevant. Not only is $N_0$ already pinned down by other means, but $\Mtrue_0(x)$, the marginal distribution of the fixed species, is available,
\item{} the species '0' is poorly known, and the data sample at hand is unable to resolve its contribution. This is the case if the~$y$ variables cannot discriminate between species '0' against any one of the other $\Ns$ species. Stated differently, if $N_0$ is left free to vary in the fit, the covariance matrix blows up for certain species and the measurement is lost. To avoid that, one is lead to accept an {\it a priori} value for $N_0$, and to compute systematics associated to the choice made for it. In that case, the worst case scenario is met if $\Mtrue_0(x)$ is unknown as well.
\end{enumerate}
It is shown below that the $\sPlot$ formalism can be extended to deal with this situation, 
whether or not $\Mtrue_0(x)$ is known, 
although in the latter case the statistical price to pay can be prohibitive.

\subsection{Assuming $\Mtrue_0$ to be known}
Here, 
it is assumed that $\Mtrue_0(x)$, 
is taken for granted. 
Then, 
it is not difficult to show that the Extended $\sPlot$, which reproduces the marginal distribution of species $\nast$, is now given by:
\begin{equation}
\label{extendedmasterequation}
N_\nast\ _s\tilde\M_\nast(\bar x)\Deltax=c_\nast\Mtrue_0(x)\Deltax+\sum_{\chiant} \sP_\nast ~,
\end{equation}
where:
\begin{itemize}
\item{}
$\sP_\nast$ is the previously defined sWeight of Eq.~(\ref{eq:weightxnotiny}):
\begin{equation}
\sP_\nast={\sum_j \V_{\nast j}\f_j\over \sum_{k} N_k\f_k + N_0 \f_0} ~,
\end{equation}
where
the covariance matrix $\V_{ij}$ is the one resulting from the fit of the $N_{i\neq 0}$
expected number of events, that is to say the inverse of the matrix:
\begin{equation}
\V^{-1}_{ij}=\sum_{e=1}^N{\f_i\f_j\over(\sum_{k} N_k\f_k + N_0 \f_0)^2} ~,
\end{equation}

\item{} $c_\nast$ is the species dependent coefficient:
\begin{equation} 
c_\nast=N_\nast-\sum_j \V_{\nast j} ~.
\end{equation}
\end{itemize}
Some remarks deserve to be made:
\begin{itemize}

\item{}
The Likelihood is now written:
\begin{equation}
\eLik=\sum_{e=1}^{N}\ln \Big\{ \sum_{i=1}^{\Ns}N_i\f_i(y_e) + N_0 \f_0(y_e) \Big\} -\Big\{ \sum_{i=1}^{\Ns}N_i + N_0 \Big\}~.
\end{equation}
Because $N_0$ is held fixed, in general, its assumed value combined with the fitted values for the $N_i$, does not maximize it:
\begin{equation}
{\partial\eLik\over\partial N_0}
=\sum_{e=1}^N{\f_0\over\sum_{k} N_k\f_k + N_0 \f_0}-1 \neq 0 ~.
\end{equation}

\item{}
It follows that the sum over the number of events per species does not equal the total number of events in the sample:
\begin{equation}
\sum_i N_i=N-N_0\left(\sum_{e=1}^N{\f_0\over\sum_{k} N_k\f_k + N_0 \f_0}\right)\neq N-N_0 ~.
\end{equation}

\item{}
Similarly, the Variance Matrix Sum Rule Eq.~(\ref{eq:VarianceSumRule}) holds only for $N_0=0$:
\begin{equation}
\sum_i N_i \V^{-1}_{ij}=1-N_0v_j ~,
\end{equation}
where the vector $v_j$ is defined by:
\begin{equation}
\label{eq:vectorvj}
v_j\equiv\sum_{e=1}^N{\f_0\f_j\over(\sum_{k} N_k\f_k + N_0 \f_0)^2} ~.
\end{equation}

\item{}
Accordingly, Eq.~(\ref{eq:CoVarianceSumRule}) becomes:
\begin{equation}
\sum_j \V_{jl}=N_l+N_0 \sum_j \V_{lj} v_j ~.
\end{equation}

\item{}
Thus, as they should, the $c_\nast$ coefficients vanish only for $N_0=0$:
\begin{equation}
c_\nast=-N_0 \sum_j \V_{nj} v_j ~.
\end{equation}

\item{}
The above defined Extended $\sPlots$ share the same properties as the $\sPlots$:

\begin{enumerate}
\item{} They reproduce the true marginal distributions, as in Eq.~(\ref{eq:sPlotsFormula}).
\item{} In particular, they are properly normalized, as in Eq.~(\ref{eq:NormalizationOK}).
\item{} The sum of $\sP_\nast^2$ reproduces $\sigma^2[N_\nast]$, as in Eq.~(\ref{eq:SumOfErrors}).
\end{enumerate}
\end{itemize}

\subsection{Assuming $\Mtrue_0$ to be unknown}
In the above treatment, because one assumes that a special species '0' enters in the sample composition, the sWeights per event do not add up to unity, as in Eq.~(\ref{eq:numberconservation}). Instead one may define the sWeights for species '0' as:
\begin{equation}
\sP_0\equiv 1-\sum_i\sP_i
\end{equation}
and introduce the reconstructed $_s\tilde\M_0$ distribution (normalized to unity):
\begin{equation}
\label{eq:motilde} 
_s\tilde\M_0(x) \Deltax~=~\left( N-\sum_{i,j}\V_{ij} \right) ^{-1}\sum_{\chiant}\sP_0 ~,
\end{equation}
which reproduces the true distribution $\Mtrue_0(x)$ if (by chance) the value
assumed for $N_0$ is the one which maximizes the Likelihood.

Taking advantage of $_s\tilde\M_0(x)$, one may redefine the Extended $\sPlots$ by:
\begin{equation}
\label{eq:revisedextendedmasterequation}
N_\nast\ _s\tilde\M_\nast(\bar x)\Deltax=c_\nast\ _s\tilde\M_0(x)\Deltax+\sum_{\chiant}  \sP_\nast
=\sum_{\chiant}  \seP_\nast ~,
\end{equation}
where the redefined sWeight which appears on the right hand side is given by:
\begin{equation}
\label{eq:redefinedsWeights}
\seP_\nast~\equiv~\sP_\nast+{N_i-\sum_j\V_{ij}\over N-\sum_{i,j}\V_{ij} }\sP_0 ~.
\end{equation}
It does not rely on {\it a priori} knowledge on the true distribution $\Mtrue_0(x)$. With this redefinition, the following properties hold:
\begin{itemize}
\item{}
The set of reconstructed $x$-distributions $N_i\tilde\M_i$ of Eq.~(\ref{eq:revisedextendedmasterequation}) completed by $(N-\sum_i N_i)\tilde\M_0$ of Eq.~(\ref{eq:motilde}) are such that they add up in each $x$-bin to the number of events observed.
\item{}
The normalization constant of the $\tilde\M_0$ distribution vanishes quadratically with $N_0$.
It can be rewritten in the form:
\begin{equation} 
N-\sum_{i,j}\V_{ij}=N_0^2\ \left(v_0 -\sum_{i,j}\V_{ij}v_iv_j\right) ~,
\end{equation}
where $v_0$ is defined as $v_j$ (cf. Eq.~(\ref{eq:vectorvj})) and where the last term is regular when $N_0\rightarrow 0$.
\item{} Whereas the normalization of the redefined extended $\sPlots$ remains correct,
the sum of the redefined sWeights Eq.~(\ref{eq:redefinedsWeights}) squared is no longer equal to $\V_{\nast\nast}$. Instead:
\begin{equation}
\sum(\seP_\nast)^2=\V_{\nast\nast}+{(N_\nast-\sum_j\V_{\nast j})^2\over N-\sum_{i,j}\V_{ij}}
=\V_{\nast\nast}+{\sum_{ij}\V_{\nast i}\V_{\nast j} v_i v_j\over v_0-\sum_{ij} \V_{ij}v_i v_j} ~.
\end{equation}
Since the expression on the right hand side is regular when $N_0\rightarrow 0$, it follows that there is a price to pay to drop the knowledge of $\Mtrue_0(x)$, even though one expects a vanishing $N_0$. Technically, this feature stems from 
\begin{equation}
\sum (\sP_0)^2=\sum \sP_0=N-\sum_{i,j}\V_{ij} ~.
\end{equation}
Hence, the sum in quadrature of the $_s\tilde\M_0$ uncertainties per bin diverges with $N_0\rightarrow 0$. This just expresses the obvious fact that no information can be extracted on species '0' from a sample which contains no such events.
\end{itemize}

\end{document}